\documentclass[12pt,prd,nofootinbib,preprint,superscriptaddress]{revtex4}
\pdfoutput=1
\usepackage{amsmath,amssymb}
\usepackage{epsfig}
\usepackage{graphicx}
\usepackage[usenames,dvipsnames]{xcolor}
\usepackage{subfigure}
\usepackage{slashed}
\usepackage{float}
\usepackage[colorlinks,citecolor=blue]{hyperref}
{}
{}

\begin{document}
	\title{\textcolor{WildStrawberry}{Connecting dark matter, baryogenesis and neutrinoless double beta decay in a $A_{4}\otimes Z_{8}$ based $\nu$2HDM}}
	
	\author{Lavina Sarma}
	\email{lavina@tezu.ernet.in}
	\affiliation{Department of Physics, Tezpur University, Tezpur 784028, India}
	
	\author{Partha Kumar Paul}
	\email{ph22resch11012@iith.ac.in}
	\affiliation{Department of Physics, Tezpur University, Tezpur 784028, India}
	\affiliation{Department of Physics, Indian Institute of Technology Hyderabad, Kandi, Sangareddy 502285, Telangana, India}
	
	\author{Mrinal Kumar Das}
	\email{mkdas@tezu.ernet.in}
	\affiliation{Department of Physics, Tezpur University, Tezpur 784028, India}
	
	\begin{abstract}
		In this paper, we discuss the impact of neutrino phenomenology and related cosmology on an $A_{4}\otimes Z_{8}$ symmetric $\nu$2HDM along with an addition of a new particle, i.e. a gauge singlet(S). The additional particle is a sterile neutrino which is considered to be a probable dark matter candidate in our work. With the choice of sterile neutrino mass in keV range we evaluate the active-DM mixing angle, decay rate and relic abundance considering various cosmological constraints. Simultaneously, a detailed analysis on baryogenesis and neutrinoless double beta decay is also carried out for low scale right-handed neutrino masses. We have considered various bounds from experiments such as Lyman-$\alpha$, X-ray observation, Planck data and KamLAND-Zen limit to validate the model w.r.t the phenomena studied in it.   \\\\


\end{abstract}
	\maketitle	
	
	\section{Introduction}
	
	We are familiar with the immense accomplishment of the Standard Model(SM) in explaining the theory for fundamental particles and its interactions. In spite of being an affluent and self-consistent model, it is certainly incomplete. Various observations point towards the need for physics beyond the Standard Model(BSM). This includes the non-zero mass of neutrinos\cite{deSalas:2017kay,Senjanovic}, Baryon Asymmetry of the Universe(BAU)\cite{leptogenesis,Hugle:2018qbw,neutrinomasspdg,Minkowski,Mohapatra,Yanagida:1979as,Schechter:1980gr,Glashow:1979nm}, Dark Matter(DM)\cite{bertone2005particle,Moore:1999nt}, etc. The neutrino oscillations\cite{neuOsc,neuOsc2} has revealed that the neutrinos are massive\cite{Lattanzi:2016rre} and also the fact that their flavors mix. The recent Neutrino experiments MINOS\cite{MINOS},RENO\cite{RENO},T2K\cite{T2K},Double-Chooz\cite{DCHOOZ} have not only confirmed but also measured the neutrino oscillation parameters more accurately\cite{Choubey}. We are well aware of the various beyond the Standard model frameworks which tends to incorporate the explanations for the above mentioned anomalies. This includes the see-saw mechanisms such as type I\cite{Minkowski}, type II\cite{ANTUSCH2004199}, type III\cite{Foot1989}, inverse see-saw\cite{HIRSCH2009454,Khalil}, radiative see-saw\cite{Ma11,Ma:2006km,Ma:2017kgb} and Minimal Extended see-saw (MES)\cite{Zhang,Das:2018qyt}. In Type I see-saw, the SM is extended by the addition of SM singlet fermions usually defined as right-handed(RH) neutrinos, that have Yukawa interactions with the SM Higgs and left-handed doublets. The light neutrino mass matrix arising from this type of see-saw is of the form $M_\nu\approx M_d M_R^{-1} M_d^T$ , where $M_d$ and $M_R$ are Dirac and Majorana masses respectively. In the type II see-saw, an additional $SU(2)_L$ triplet scalar field is introduced. The type III see-saw introduces an additional fermion triplet field. Inverse see-saw requires the existence of extra singlet fermion to provide rich neutrino phenomenology. In this formalism,the lightest neutrino mass matrix is given by $M_\nu\approx M_d (M^T)^{-1}\mu M^{-1} M_d^T$, where $M_d$ is the Dirac mass term and $\mu$ is the Majorana mass term for sterile, while M represents the lepton number conserving interaction between right handed and sterile fermions. In the Minimal Extended see-saw (MES), which is an extension of the canonical type-I see-saw, three additional right handed neutrinos and one gauge singlet chiral fermion field S as a sterile neutrino are included to the standard model particles. In this formalism, the $4\times4$ active-sterile neutrino mass matrix is given by
	\begin{equation}
	M_\nu^{4\times4}=-\begin{pmatrix}
	M_D M_R^{-1}M_D^T & M_D M_R^{-1} M_S^T\\
	M_S(M_R^{-1})^TM_D^T & M_S M_R^{-1} M_S^T
	\end{pmatrix}\label{M4}
	\end{equation}
	where $M_D$, $M_R$ and $M_S$ are the Dirac, Majorana and Sterile neutrino mass matrices. Now the light neutrino mass matrix can be written as, \cite{Zhang}
	\begin{equation}
	M_\nu^{3\times3}\simeq M_D M_R^{-1}M_S^T(M_SM_R^{-1}M_S^T)^{-1}M_S(M_R^{-1})^TM_D^T-M_DM_R^{-1}M_D^T
	\end{equation}
	and the sterile neutrino mass as,
	\begin{equation}\label{eq:6}
	m_s\simeq -M_S M_R^{-1}M_S^T
	\end{equation} Neutrinos with sub-eV scale are obtained from $M_D$ at electroweak scale, $M_R$ at TeV scale and $M_S$ at keV scale.\\
	
	Also there are significant lines of evidence of DM which comprises of observations in galaxy cluster by Fritz Zwicky \cite{Zwicky:1933gu} in 1933, gravitational lensing (which could allow galaxy cluster to act as gravitational lenses as postulated by Zwicky in 1937) \cite{Treu:2012sn}, galaxy rotation curves in 1970 \cite{Rubin:1970zza}, cosmic microwave background \cite{cosmicmicrowave} and the most recent cosmology data given by Planck satellite \cite{Ade:2015fva}. From the recent Planck satellite data, it is certain that approximately $27\%$ of the present Universe is comprised of DM, which is about five times more than the baryonic matter. The present dark matter abundance is reported as
	\begin{equation*}
	\Omega_{DM}h^2 = 0.1187 \pm 0.0017.
	\end{equation*}
	Searching for the possible DM candidates with new physics beyond standard model has been a great challenge to the physics community world wide. The important criteria to be fulfilled by a particle to be considered as a good DM candidate can be found in \cite{Taoso:2007qk}. These requirements exclude all the SM particles from being DM candidate. This has motivated the particle physics community to study different possible BSM frameworks which can give rise to the correct DM phenomenology and can also be tested at several experiments.\\
	The baryon asymmetry of the Universe, also known as the baryogenesis, is one of the most unperceived problems of Particle Physics as well as Cosmology, which is the observed imbalance in the baryonic matter and anti-baryonic matter in the observable Universe. The three Shakharov conditions have to be satisfied for the particle to obtain a significant amount of baryogenesis, which demands baryon number (B) violation, C and CP violation and departure from thermal equilibrium. As none of the SM particles full fill these conditions in an adequate amount, we need to go beyond the SM. We can incorporate such a mechanism via leptogenesis. The leptogenesis or lepton asymmetry is generated by the out-of-equilibrium CP-violating decays of right hand neutrino which can be converted into baryon asymmetry through a process called the sphaleron process. Together with the cosmic baryon asymmetry of the Universe, we have explicit reasons to extend the SM with new particles and fields.\\
	Motivated by these factors, we have done a phenomenological study on the neutrino two Higgs doublet model ($\nu$2HDM) in the framework of Minimal Extended see-saw in which the particle content of SM is extended by a new Higgs doublet field ($\eta$), three singlet neutral fermions ($N_i$) and one sterile singlet neutrino (S). We have constructed the Dirac mass matrix in such a way that the $\mu-\tau$ symmetry is broken to generate the non-zero reactor mixing angles. We calculate the effective neutrino mass using the global fit analysis data in the 3$\sigma$ range and studied its variation with lightest neutrino mass and compared the value of effective mass with KamLAND-Zen bounds both in NH and IH cases. 
	We have studied the DM phenomenology considering the
	lightest sterile neutrino(S) as a potential DM candidate. We have
	evaluated the model parameters and then calculated DM
	mass, DM active mixing, relic density and the decay rate of the sterile neutrino. We then calculate the asymmetry in lepton flavor in the decay process of the lightest right handed neutrino. Our model is satisfying the limits obtained from cosmology as well as astrophysics.
	\hspace{0.5cm}\\
	This paper is organized as follows. In section \ref{sec2} we have discussed the generic $\nu$2HDM following the $A_4\times Z_8$ model and generation of the mass matrices in the leptonic sector. The section \ref{sec3} discusses the sterile dark matter, X-ray and Ly-$\alpha$ constraints. Section \ref{sec4} is the discussion of leptogenesis. Numerical analysis and results are discussed in section \ref{sec5}. Finally, the summary of our work is concluded in the section \ref{sec6}.

	\section{$A_{4}\otimes Z_{8}$ flavor symmetric Neutrino Two Higgs Doublet Model}\label{sec2}
	
	It will not be natural if we consider that the neutrino mass comes from the SM Higgs doublet (H) as the neutrino Yukawa coupling constant is too small compared to other leptons and quarks. This problem can be solved if we consider that the neutrino mass comes from another scalar doublet with naturally small vacuum expectation value (vev).
	$\nu$2HDM model \cite{nu2HDM} is one of the natural choices for beyond-SM models containing two Higgs doublets instead of just one.\\
	Consider the minimal Standard Model with three lepton families:\\
	\begin{center}$\begin{pmatrix}
		\nu_i\\
		l_i\\
		\end{pmatrix}_L\sim(1,2,-1/2), l_{iR}\sim(1,1,-1)$
	\end{center}
	Now we add three neutral fermion singlets, $N_{iR}\sim(1,1,0)$. We assign them L=0 instead of L=1 to forbid the Yukawa coupling term with SM Higgs doublet. Now we introduce a new scalar doublet $\eta$ as,
	\begin{equation*}
	\begin{pmatrix}
	\eta^+\\
	\eta^0\\
	\end{pmatrix}\sim(1,2,1/2)
	\end{equation*} with L= -1.\\
	The new Yukawa interaction and mass terms are
	\begin{equation}
	-\mathcal{L}_Y\sim y\bar{L}\tilde{\eta} N+\frac{1}{2}\bar{N^c}m_NN+h.c.
	\end{equation}
	where $\tilde{\eta}=i\sigma_2\eta^*$\\
	Here, we see that the lepton number is violated by 2 units in the first term of the Yukawa interaction Lagrangian.
	Similar to the type-I seesaw, the mass matrix for light neutrinos can be written as:
	\begin{equation}
	M_\nu=M_D M_R^{-1}M_D^T
	\end{equation}
	where, $m_D=<\tilde{\eta}>y$ and $<\tilde{\eta}>=v$.\\
	In our work, we extend the $\nu$2HDM\cite{nu2HDM,Sarma:2021icl} by a gauge singlet fermion(S). The main motivation of this extended field is to incorporate dark matter phenomenology in our model. Furthermore, a flavor symmetric realization of this extension of $\nu$2HDM is done with the help discrete flavor symmetries $A_{4}$ and $Z_{8}$. Non-Abelian discrete flavor symmetries play important role in model building \cite{Nonabelian,nu2HDM,King}. $A_4$ being the discrete symmetry group of rotation leaving a tetrahedron
	invariant. It has 12 elements and 4 irreducible representation denoted by 1,$1^\prime$,$1^{\prime\prime}$ and 3.\\
	Along with the particle content of our model in addition to the SM particles (i.e three right handed neutrinos ($N_{1}$, $N_{2}$, $N_{3}$), one Higgs doublet ($\eta$) and one additional gauge singlet ($S$), we also introduce three sets of flavon fields $\varphi$, $\xi$ and $\chi$. Here, we have assigned left-handed lepton doublet ($\ell$) to transform as $A_4$ triplet whereas right-handed charged leptons ($e_R,\mu_R,\tau_R$) transform as 1, $1^{\prime\prime}$ and $1^\prime$ respectively. An extra discrete symmetry $Z_8$ has been introduced in order to distinguish the neutrino and the flavon fields.\\
	The particle content and charge assignments are shown in Table \ref{TAB1}.
	
	\begin{table}[h]
		
		\centering
		\begin{tabular}{c c c c c c c c c c c c c c c c c}
			\hline\hline
			Field & $\ell$ & $e_R$ & $\mu_R$ & $\tau_R$ & $H$ & $\eta$ & $\varphi$ & $\varphi^\prime$ & $\varphi^{\prime\prime}$ & $\xi$ & $\xi^\prime$ & $\chi$ & $N_{1}$ & $N_{2}$ & $N_{3}$ & $S$ \vspace{1.5mm}\\  
			SU(2) & 2 & 1 & 1 & 1 & 2 & 2 & 1 & 1 & 1 & 1 & 1 & 1 & 1 & 1 & 1 &1 \vspace{1.5mm}\\ 
			$A_4$ & 3 & 1 & $1^{\prime\prime}$ & $1^\prime$ & 1 & 1 & 3 & 3 & 3 & 1 & $1^\prime$ & 1& 1 & $1^\prime$ & 1 &1 \vspace{1.5mm}\\ 
			$Z_8$ &$\omega^3$ & $\omega^4$&$\omega^4$&$\omega^4$&$\omega^2$&$\omega$&$\omega^7$&$\omega^2$&$\omega^3$&$\omega^6$&$\omega^4$&$\omega$&$\omega^5$&$\omega^2$&$\omega$&$\omega^2$\\
			\hline\hline
		\end{tabular} 
		\caption{Particle content and their charge assignments under SU(2), $A_4$ and $Z_8$ group.}
		\label{TAB1}
	\end{table}
	
	\begin{table}[h]
		
		\centering
		\begin{tabular}{c c c c }
			\hline\hline
			Field & $\zeta$ & $\zeta^\prime$ & $\zeta^{\prime\prime}$ \vspace{1.5mm}\\ 
			$A_4$ & 3 & 3 & 3  \vspace{1.5mm}\\ 
			$Z_8$ &$\omega^7$ &$\omega^2$ &$\omega^3$ \\
			\hline\hline
		\end{tabular} \label{TAB4}
		\caption{Charge assignment of singlet flavons under $A_4$ and $Z_8$.}
		
	\end{table}
	
	The leading order invariant Yukawa Lagrangian for the lepton sector is given by,
	\begin{equation}
	\mathcal{L}= \mathcal{L}_{M_l}+ \mathcal{L}_{M_D^\prime}+ \mathcal{L}_{M_R}+ \mathcal{L}_{M_S}+h.c. 
	\end{equation}
	where,
	\begin{equation}
	\mathcal{L}_{M_l}=\frac{y_e}{\Lambda}(\bar{l}H\varphi)_{\underline{1}}e_R+\frac{y_\mu}{\Lambda}(\bar{l}H\varphi)_{\underline{1^\prime}}\mu_R+\frac{y_\tau}{\Lambda}(\bar{l}H\varphi)_{\underline{1^{\prime\prime}}}\tau_R 
	\end{equation}
	\begin{equation}
	\mathcal{L}_{M_D^{\prime}}=\frac{y_1}{\Lambda}(\bar{l}\tilde{\eta}\varphi)_{\underline{1}}N_{1}+\frac{y_2}{\Lambda}(\bar{l}\tilde{\eta}\varphi^\prime)_{\underline{1^{\prime\prime}}}N_{2}+\frac{y_3}{\Lambda}(\bar{l}\tilde{\eta}\varphi^{\prime\prime})_{\underline{1}}N_{3}
	\end{equation}
	\begin{equation}
	\mathcal{L}_{M_R}=\frac{1}{2}\lambda_1\xi\bar{N^c_{1}}N_{1}+\frac{1}{2}\lambda_2\xi^\prime\bar{N^c_{2}}N_{2}+\frac{1}{2}\lambda_3\xi\bar{N^c_{3}}N_{3}  
	\end{equation}
	\begin{equation}
	\mathcal{L}_{M_S}=\frac{1}{2}\rho\chi \bar{S^c} N_{1}
	\end{equation}
	where $\Lambda$ denotes the cut-off scale. To generate the desired light neutrino mas matrix we choose the vev alignments of the extra flavons as following,
	\begin{center}
		$<\varphi>=(\kappa,0,0)$ , $<\varphi^\prime>$=$<\varphi^{\prime\prime}>=(\kappa,\kappa,\kappa)$ , $<\xi>=<\xi^\prime>=\kappa$ , $<\chi>=u$
	\end{center}
	Following the $A_4$ product rules \cite{Nonabelian} and using the above mentioned vev alignment, we can obtain
	the charged lepton mass matrix as follows
	\begin{equation}
	M_l=\frac{<H>\kappa}{\Lambda}\begin{pmatrix}
	y_e & 0 & 0 \\
	0 & y_\mu & 0 \\
	0 & 0 & y_\tau \\
	\end{pmatrix}
	\end{equation}
	The Dirac mass matrix is given by,
	\begin{equation}
	M_D^\prime=\begin{pmatrix}
	a & b & c \\
	0 & b & c \\
	0 & b & c \\
	\end{pmatrix}
	\end{equation}
	The right handed neutrino mass matrix is,
	\begin{equation}
	M_R=\begin{pmatrix}
	M_1 & 0 & 0 \\
	0 & M_2 & 0 \\
	0 & 0 & M_3 \\
	\end{pmatrix}.
	\end{equation}
	And the sterile neutrino mass matrix $M_S$ can be written as,
	\begin{equation}
	M_S=\begin{pmatrix}
	s & 0 & 0
	\end{pmatrix}
	\end{equation}
	where, $a=\frac{<\tilde{\eta}>\kappa}{\Lambda} y_1$, $b=\frac{<\tilde{\eta}> \kappa}{\Lambda} y_2$, $c=\frac{<\tilde{\eta}> \kappa}{\Lambda} y_3$, $M_1$=$\lambda_1\kappa$, $M_2$=$\lambda_2\kappa$, $M_3$=$\lambda_3\kappa$ and $s=\rho u$.\\
	Considering these $M_D^\prime$, $M_R$ and $M_S$ in the light neutrino mass matrix\cite{Zhang}, we get; \begin{equation}
	M_\nu^{3\times3}\simeq M_D^{\prime} M_R^{-1}M_S^T(M_SM_R^{-1}M_S^T)^{-1}M_S(M_R^{-1})^TM_D^{\prime T}-M_D^{\prime}M_R^{-1}M_D^{\prime T}
	\end{equation}
	\begin{equation}
	M_\nu^{3\times3}=\begin{pmatrix}
	-\frac{b^2}{M_2}-\frac{c^2}{M_3} & -\frac{b^2}{M_2}-\frac{c^2}{M_3} & -\frac{b^2}{M_2}-\frac{c^2}{M_3}\vspace{2mm}\\
	-\frac{b^2}{M_2}-\frac{c^2}{M_3} & -\frac{b^2}{M_2}-\frac{c^2}{M_3} & -\frac{b^2}{M_2}-\frac{c^2}{M_3}\vspace{2mm}\\ -\frac{b^2}{M_2}-\frac{c^2}{M_3} & -\frac{b^2}{M_2}-\frac{c^2}{M_3} & -\frac{b^2}{M_2}-\frac{c^2}{M_3}\\
	\end{pmatrix}
	\end{equation}
	It is clear that this matrix is a symmetric matrix generated by $M_D^\prime$, $M_R$ and $M_S$ matrices. It can produce only one mixing angle and one mass square difference. This symmetry must be broken in order to generate two mass square differences and three mixing angles. In
	order to introduce $\mu-\tau$ asymmetry in the light neutrino mass matrix we introduce $SU(2)_{L}$ singlet flavon fields $\zeta$, $\zeta^{'}$ and $\zeta^{''}$ which breaks the $\mu$-$\tau$ symmetric after being incorporated in the Dirac mass matrix ($M_D^{\prime}$). This additional matrix has a crucial role to play in reproducing non-zero reactor mixing angle. And the Lagrangian responsible for generating the matrix can be written as,
	\begin{equation}
	\mathcal{L}_{M_P}=\frac{y_1}{\Lambda}(\bar{l}\tilde{\eta}\zeta)_{\underline{1}}N_{1}+\frac{y_2}{\Lambda}(\bar{l}\tilde{\eta}\zeta^\prime)_{\underline{1^{\prime\prime}}}N_{2}+\frac{y_3}{\Lambda}(\bar{l}\tilde{\eta}\zeta^{\prime\prime})_{\underline{1}}N_{3}.
	\end{equation}
	Taking vev alignment for the new flavon field as: $<\zeta>$=$<\zeta^\prime>$=$<\zeta^{\prime^\prime}>=(0,\kappa,0)$, we get the matrix
	as,
	\begin{equation}
	M_P=\begin{pmatrix}
	0 & 0 & p \\
	0 & p & 0 \\
	p & 0 & 0 \\
	\end{pmatrix}.
	\end{equation}
	Hence $M_D$ takes the new form as,
	\begin{equation}
	M_D=M_D^\prime+M_P=\begin{pmatrix}
	a & b & c+p \\
	0 & b+p & c \\
	p & b & c \\
	\end{pmatrix}.
	\end{equation}
	Therefore, we modify the light neutrino mass matrix by replacing $M_{D}^{\prime}$ by $M_{D}$ which results in
	\begin{equation}
	M_\nu^{3\times3}\simeq M_D M_R^{-1}M_S^T(M_SM_R^{-1}M_S^T)^{-1}M_S(M_R^{-1})^TM_D^T-M_DM_R^{-1}M_D^T.
	\end{equation}
	The final light neutrino mass matrix is thus given by:
	\begin{equation}
	M_\nu^{3\times3}=\begin{pmatrix}
	-\frac{b^2}{M_2}-\frac{(c+p)^2}{M_3} & -\frac{-b(b+p)}{M_2}-\frac{c(c+p)}{M_3} & -\frac{b^2}{M_2}-\frac{c(c+p)}{M_3}\vspace{2mm}\\
	-\frac{-b(b+p)}{M_2}-\frac{c(c+p)}{M_3} & -\frac{c^2}{M_2}-\frac{(b+p)^2}{M_3} & -\frac{c^2}{M_2}-\frac{b(b+p)}{M_3}\vspace{2mm}\\ -\frac{b^2}{M_2}-\frac{c(c+p)}{M_3} & -\frac{c^2}{M_2}-\frac{b(b+p)}{M_3} & -\frac{b^2}{M_2}-\frac{c^2}{M_3}\\
	\end{pmatrix}.
	\end{equation}
	The $4\times4$ active sterile neutrino mass matrix represented as in Eq.\ref{M4} becomes,
	\begin{equation}
	M_\nu^{4\times4}=\begin{pmatrix}
	-\frac{a^2}{M_1}-\frac{b^2}{M_2}-\frac{(c+p)^2}{M_3} & -\frac{b(b+p)}{M_2}-\frac{c(c+p)}{M_3} & -\frac{b^2}{M_2}-\frac{a p}{M_1}-\frac{c(c+p)}{M_3} & -\frac{a s}{M_1}\vspace{2mm}\\
	-\frac{b(b+p)}{M_2}-\frac{c(c+p)}{M_3} & -\frac{c^2}{M_3}-\frac{(b+p)^2}{M_2} & -\frac{c^2}{M_3}-\frac{b(b+p)}{M_2} & 0\vspace{2mm}\\ -\frac{b^2}{M_2}-\frac{a p}{M_1}-\frac{c(c+p)}{M_3} & -\frac{c^2}{M_3}-\frac{b(b+p)}{M_2} & -\frac{b^2}{M_2}-\frac{c^2}{M_3}-\frac{p^2}{M_1}&-\frac{p s}{M_1} \vspace{2mm}\\
	-\frac{a s}{M_1} &0 & -\frac{p s}{M_1} & \frac{s^2}{M_1}
	\end{pmatrix}.
	\end{equation}
	In $M_\nu^{4\times4}$, there exists three eigenstates for three active neutrinos and one for the light sterile neutrino.\\
	Since we have included one extra generation of neutrino along with the active neutrinos in our model thus, the final neutrino mixing matrix for the active-sterile mixing takes 4 $\times$ 4 form as \cite{Zhang},
	\begin{equation}
	V\simeq \begin{pmatrix}
	(1-\frac{1}{2}RR^\dagger)U_{PMNS} & R \\
	-R^\dagger U_{PMNS} & 1-\frac{1}{2}R^\dagger R\\
	\end{pmatrix}\label{M6}
	\end{equation}
	where $R=M_D M_R^{-1} M_S^T(M_SM_R^{-1}M_S^T)^{-1}$ is a 3$\times$1 matrix representing the strength of active sterile mixing and $U_{PMNS}$ is the leptonic mass matrix for active neutrinos given by Eq. \ref{M5}.\\
	
	\section{Sterile Dark Matter}\label{sec3}
		A minimally extended Standard Model of particle physics can easily accommodate a non resonantly produced sterile dark matter(DM). With the mass of sterile neutrino restricted to a few keV, it can be considered as a prime candidate for warm DM. The sterile neutrinos couples with the standard model particles only via its mixing to the active neutrinos. Thus, from the active neutrinos which are a part of the primordial plasma (as they are weakly interacting), the DM abundance can be eventually build up. This mechanism is also known as the Dodelson and Widrow (DW) mechanism. The non resonant production (NRP) of sterile neutrinos takes place in absence of lepton asymmetry. However, for compelling amount of lepton asymmetry in the primordial plasma, we can get resonant sterile neutrinos having very small mixing angles and thereby producing notable colder momenta. This production mechanism is also known as Shi $\&$ Fuller(SF) or resonant production mechanism\cite{DM1}. A minimum amount of dark matter contribution which can be produced as a consequence of the dark matter mass and the mixing angle is accounted by non resonant production mechanism.
	In our model, we have considered a singlet sterile fermion($S$) which acts as a non resonant DM candidate. Thus, from now on we denote the sterile fermion mass as $m_{DM}$. We solve the model parameters with some fixed values of variables such as $ M_{1}$, $ M_{2}$ and $ M_{3}$ in the range $10^{4}-10^{5}$ GeV, $5\times10^{5}- 10^{6}$ GeV and $10^{7}-10^{8}$ GeV respectively. Also the lightest of the active neutrinos is considered very small having mass in the range $10^{-10}-10^{-9}$ eV. On computing these values we are able to get the desired mass $m_{DM}$, i.e in keV range and also the mixing angle $ sin^{2}(2\theta_{DM})$ satisfying the cosmological bounds. In our work, the contribution towards the mixing angles comes from $V_{14}$ and $V_{34}$, non-vanishing components of the mixing matrix V (\ref{M6}).
		The relic abundance of any species can be expressed as \cite{kolb2018early},
		\begin{equation}
		\Omega_{x}h^2=\frac{\rho_{x_{0}}}{\rho_{crit}}=\frac{s_0 Y_\infty m}{\rho_{crit}}
		\end{equation}
		where, $\rho_{x_{0}}$ is present energy density of $x$, $\rho_{crit}$ represents the critical energy density of the universe, $s_0$ is the present day entropy and $Y_\infty$ is the present abundance of the particle $x$. Also we can get the values of $\rho_{crit}\approx 1.054*10^{-5} h^2~\rm GeV~ cm^{-3}$ and $s_0 \approx 2886~\rm cm^{-3}$ from Particle Data Group(PDG).\\
		Further in case of sterile neutrinos, it can be expressed by the relation: \cite{relic}
		\begin{equation}
		\Omega_{\alpha x}=\frac{m_x Y_{\alpha x}}{3.65\times10^{-9} h^2~\rm GeV}.
		\end{equation}
		where $\alpha= e, \mu, \tau$.\\
		Now the resulting relic abundance of any sterile neutrino state with a non-vanishing mixing to the active neutrinos, is proportional to the active-sterile mixing and the mass of the sterile, which is again expressed as\cite{abada2014dark,ng2019new},
		\begin{equation}
		\Omega_{\alpha S} h^2=1.1\times10^7\sum C_\alpha(m_s)|V_{\alpha s}|^2(\frac{m_s}{\rm keV})^2
		\end{equation}
		where, 
		\begin{equation}
		C_\alpha(m_s)=2.49\times10^{-5}\frac{Y_{\alpha s}~\rm keV}{sin^2(\theta_{\alpha s}) m_s}.
		\end{equation}
		$C_\alpha$ are active flavor dependent coefficients and can be numerically computed by solving Boltzmann equation\cite{relic1}.\\
		Using the parametrization $|V_{\alpha s}\simeq sin(\theta_{\alpha s})|$
		and with the consideration of sterile neutrino as a dark matter candidate, we replace the symbol $s$ by $DM$ in the following expression for relic abundance. Therefore, the simplified equation for relic abundance for non resonantly produced dark matter takes the form \cite{abazajian2001sterile, relic1, relic, Abada}:
		\begin{equation}
		\Omega_{ DM} h^{2}\simeq 0.3\times 10^{10} sin^{2}(2\theta_{DM})\big(\frac{m_{DM}\times 10^{-2}}{\rm keV}\big)^{2}
		\end{equation} 
		where, $\Omega_{DM}$ is directly proportional to $m_{DM}$ which is the DM mass as mentioned earlier and 
		$sin^{2}(2\theta_{DM})$ viz the active-DM mixing angle with $sin^2(2\theta_{DM})$ = $4(V^2_{14}+V^2_{34})$.
	\subsection{Constraints on non resonant sterile dark matter }
	
	The decay rate of a sterile neutrino when it radiatively decays into an active neutrino and a photon $\gamma$ with its energy $E_{\gamma}=\frac{m_{DM}}{2}$ is strictly dependent on dark matter mass. Thus, sterile neutrinos bearing masses above the keV range is significantly ruled out under the above mentioned condition. The decay rate for the process $S \longrightarrow \nu + \gamma $ is given by\cite{DM12}:
	\begin{equation}
	\Gamma\simeq 1.38\times 10^{-22} sin^{2}(2\theta_{DM})\big(\frac{m_{DM}}{\rm keV}\big)^{5} s^{-1}.
	\end{equation}  
	For a particle to be a DM candidate, one of the crucial conditions is its stability. Thus, despite the probable decay of the sterile neutrino, it can be considered as a DM candidate due to the negligible decay rate of $S$ as a consequence of the mixing angles.
	
	A significant experimental observation which can be considered for analyzing cosmology is the Lyman-$\alpha$ observations. Based on this Lyman-$\alpha$ observation, various studies over the past decade have put forward constraints on the non-resonant sterile neutrino DM. The constraints obtained are applicable to thermal relic WDM  as they are equivalent to non-resonant sterile DM and thus, we can find a conversion relation between non-resonant sterile DM mass($ m_{nrs}$) and thermal relic WDM mass($ m_{WDM}$)\cite{DM2}. Thermal relic masses below 0.75keV (approx) was ruled out earlier in\cite{DM15}, however, later a more stringent bound was found by\cite{DM16} of $ m_{WDM}$ $\geq$ 2.5 keV. The latter constraint was based on the SDSS data, which permitted to exclude non-resonantly produced sterile neutrinos as probable DM candidate. Again, a few years back a 2-$\sigma$ limit on $ m_{WDM} \geq 3.3$ keV corresponding to $ m_{nrs} \geq 18.5$ keV was recounted by\cite{dm18}. Recent works could give more tighter bounds as they used more than 13000 quasar spectra from the BOSS survey. It therefore, reported a 2-$\sigma$ limit of $ m_{WDM}\geq 4.35$ keV (corresponding to $ m_{nrs}\geq 26.4$ keV\cite{DM19}).
	The analysis done by\cite{DM2} shows that the Milky-way satellite counts discards models wherein the particle mass is below $ m_{DM}\sim$ 4.5 keV and larger region of higher order masses. This allows the parameter space near around the line signal, i.e $ m_{DM}$=7.1 keV. Again, from V13 model based Lyman-$\alpha$ limit and X-ray bounds, a narrow area above $ m_{DM}\sim$ 10 keV remains allowed. Interestingly, the generalized Lyman-$\alpha$ bound coming from B15 reference model thoroughly overlaps with the Suzaku X-ray limits , thereby, disfavoring the entire regime of resonantly produced sterile neutrino DM. Thus, a very significantly robust limit is drawn for both non-resonant and resonant production of sterile neutrino.
	
	\section{LEPTOGENESIS}\label{sec4}
	We study baryogenesis in neutrino two Higgs doublet model realized by $A_{4} \times Z_{8}$ symmetry. Leptogenesis\cite{leptogenesis} is the process by which the observed baryon asymmetry of the Universe(BAU) can be produced. In the beyond standard model, the right handed neutrino with the lightest
	mass decays to a Higgs doublet and a lepton doublet. The Majorana property of the neutrino produces a significant amount of lepton asymmetry in the decay process. The standard model Lagrangian conserves both lepton number and baryon numbers. But chiral anomalies in BSM theories results in violation in (B+L) while conserving (B-L). The lepton asymmetry is generated by the out-of-
	equilibrium CP-violating decays of right hand neutrino, in our case $N_{1}$. As discussed in many literatures\cite{Davidson:2002qv,Buchmuller:2002rq}, we now know that there exists a lower bound of about 10TeV for the lightest of the RHNs($M_{1}$) considering the vanilla leptogenesis scenario\cite{Hugle:2018qbw,Borah:2018rca}. For a hierarchical mass of RHN, i.e $M_{1} << M_{2}, M_{3}$, the leptogenesis produced by the decay of $N_{2}$ and $N_{3}$ are suppressed due to the strong washout effects produced by $N_{1}$ or $N_{2}$ and $N_{3}$ mediated interactions\cite{Borah:2018rca}. Thereby, the lepton asymmetry is produced only by the virtue of $N_{1}$ decay and the lepton asymmetry can be converted to baryon asymmetry via violation of (B+L) through
	the process known as \textquoteleft sphalerons\textquoteright\cite{Dine:2003ax}.
	In the calculation of leptogenesis, one important quantity that differentiates between weak and strong washout regime is the decay parameter. It is expressed as:
	\begin{equation}
	K_{1}= \frac{\Gamma_{1}}{H(z=1)},
	\end{equation}
	where, $\Gamma_{1}$ gives us the total $N_{1}$ decay width, $H$ is the Hubble parameter and $z= \frac{M_{1}}{T}$ with $T$ being the temperature of the photon bath. We can express $H$ in terms of $T$ and the corresponding equation is given by:
	\begin{equation}\label{eq:1}
	H = \sqrt\frac{8\pi^{3}g_{*}}{90}\dfrac{T^{2}}{M_{Pl}}.
	\end{equation} 
	In Eq.\eqref{eq:1}, $g_{*}$ stands for the effective number of relativistic degrees of freedom and $M_{Pl}\simeq 1.22\times 10^{19}$ GeV is the Planck mass. We consider the RHN masses as $M_{1}=10^{4}-10^{5}$ GeV, $M_{2}=5\times10^{5}-10^{6}$ GeV and $M_{3}=10^{7}-10^{8}$ GeV. Now, by this choice of RHN masses along with $m_{\eta^{0}_{R}}= 1-10$ GeV and most significantly the lightest active neutrino mass $m_{l}= 10^{-10}-10^{-9}$ eV, we fall on the weak washout regime. 
	The Yukawa couplings obtained by solving the model parameters are incorporated in the decay rate equation for $N_{1}$ which is given by,
	\begin{equation}
	\Gamma_{1}= \frac{M_{1}}{8\pi}(Y^{\dagger}Y)_{11}\left[1- \Big(\frac{m_{\eta^{0}_{R}}}{M_{1}}\Big)^{2}\right]^{2}= \frac{M_{1}}{8\pi}(Y^{\dagger}Y)_{11}(1-\eta_{1})^{2}
	\end{equation}
	Again for the decays $N_{1}\rightarrow l\eta, \bar{l}\eta^{*}$, the CP asymmetry parameter $\epsilon_{1}$ is given by, 
	\begin{equation}
	\epsilon_{1}= \frac{1}{8\pi(Y^{\dagger}Y)_{11}}\sum_{j\ne 1}Im[(Y^{\dagger}Y)^{2}]_{1j}\left[ f(r_{j1},\eta_{1})- \frac{\sqrt{r_{j1}}}{r_{j1}-1}(1-\eta_{1})^{2}\right],
	\end{equation}{\label{bau}}
	where,
	\begin{equation}
	f(r_{j1},\eta_{1})= \sqrt{r_{j1}}\left[1+ \frac{(1-2\eta_{1}+r_{j1})}{(1-\eta_{1})^{2}}  ln(\frac{r_{j1}-\eta_{1}^{2}}{1-2\eta_{1}+r_{j1}})\right],
	\end{equation}
	and
	$r_{j1}= \big(\frac{M_{j}}{M_{1}}\big)^{2}$, $\eta_{1}\equiv \big(\frac{m_{\eta^{0}_{R}}}{M_{1}}\big)^{2}$.\\
	The Yukawa couplings participating in Eq.\eqref{bau} are obtained from the model on solving the model parameters.
	The Boltzmann equations for the number densities of $N_{1}$ and $N_{B-L}$, given by \cite{Davidson:2002qv},
	\begin{equation}\label{eq:3}
	\frac{dn_{N_{1}}}{dz}= -D_{1}(n_{N_{1}} - n_{N_{1}}^{eq}),
	\end{equation}
	\begin{equation}\label{eq:4}
	\frac{dn_{B-L}}{dz}= -\epsilon_{1}D_{1}(n_{N_{1}} - n_{N_{1}}^{eq})- W_{1}n_{B-L},
	\end{equation}
	respectively. In Eq.\ref{eq:3} and \ref{eq:4}, the terms have their usual meaning as discussed in many literatures\cite{Borah:2018rca,Lavina}.  
	The final $B-L$ asymmetry $n_{B-L}^{f}$ is evaluated by numerically calculating Eq.\eqref{eq:3} and Eq.\eqref{eq:4} before the sphaleron freeze-out. This is converted into the baryon-to-photon ratio given by:
	\begin{equation}\label{eq:5}
	n_{B}= \frac{3}{4}\frac{g_{*}^{0}}{g_{*}}a_{sph}n_{B-L}^{f}\simeq 9.6\times 10^{-3}n_{B-L}^{f},
	\end{equation} 
	In Eq.\eqref{eq:5}, $g_{*}= 106.75$ is the effective relativistic degrees of freedom at the time when final lepton asymmetry was produced, $g_{*}^{0}= \frac{43}{11}$ is the effective degrees of freedom at the recombination epoch and $a_{sph}=\frac{8}{23}$ is the sphaleron conversion factor taking two Higgs doublet into consideration. On solving the model parameters and incorporating in the equations responsible for generating BAU as mentioned in this section, we get quite satisfactory results abiding the Planck limit viz  $(6.04\pm0.08)\times 10^{-10}$ \cite{Aghanim:2018eyx} .

	\section{Numerical analysis and results}\label{sec5}
	The leptonic mixing matrix for active neutrinos depends on three mixing angles $\theta_{13}$, $\theta_{23}$ and $\theta_{12}$ and one CP-violating phase ($\delta$) for Dirac neutrinos and two Majorana phases $\phi_1$ and $\phi_2$ for Majorana neutrino. The Leptonic mass matrix for active neutrino is parameterized as,
	
	\begin{equation}
	U_{PMNS}=\begin{pmatrix}
	c_{12}c_{13} & s_{12}c_{13} & s_{13}e^{-i\delta} \\
	-s_{12}c_{23}-c_{12}s_{23}s_{13}e^{i\delta} & c_{12}c_{23}-s_{12}s_{23}s_{13}e^{i\delta} & s_{23}c_{13} \\
	s_{12}s_{23}-c_{12}c_{23}s_{13}e^{i\delta} & -c_{12}s_{23}-s_{12}c_{23}s_{13}e^{i\delta} & c_{23}c_{13} \\
	\end{pmatrix}.P \label{M5}
	\end{equation}
	Here,  $c_{ij}=cos\theta_{ij}$, $s_{ij}=sin\theta_{ij}$ and $P$ would be a unit matrix in the Dirac case but in Majorana case $P= diag(e^{\frac{i\phi_1}{2}},1,e^{\frac{i\phi_2}{2}}$)\\
	The light neutrino mass matrix  is diagonalized by the unitary PMNS matrix as,
	\begin{equation}
	M_\nu^{3\times3}=U_{PMNS}\cdot diag(m_1,m_2,m_3) \cdot  U_{PMNS}^T
	\end{equation}
	where $m_1$, $m_2$ and $m_3$ are three active neutrino masses.\\
	Now the active-sterile neutrino mass matrix can be diagonalized by the mixing matrix, V as,
	\begin{equation}\label{eq:7}
	M_\nu^{4\times4}=V\cdot diag(m_1,m_2,m_3,m_4) \cdot V^T
	\end{equation}
	The sterile mass is obtained from Eq.\eqref{eq:6} as:
	\begin{equation}
	m_s=\frac{s^2}{M_1},
	\end{equation}
	which is considered as the dark matter mass and is represented as $m_{DM}$ in our analysis.

	\begin{table}
		\begin{center}
			\begin{tabular} { |c|c| }
				\hline
				Free Parameter & NH/IH \\
				\hline
				$M_1(\rm GeV)$ & $10^4$ - $10^5$\\
				\hline
				$M_2(\rm GeV)$ & $5\times10^5$ - $10^{6}$ \\
				\hline
				$M_3(\rm GeV)$ & $10^{7}$ - $10^{8}$ \\
				\hline
				$m_l(\rm eV)$ & $10^{-10}$ - $10^{-9}$ \\
				\hline
				$s(\rm GeV)$ & $0.1$ - $1$  \\
				\hline
			\end{tabular}
			
			\caption{Range of the free parameters.}\label{TAB2}
		\end{center}
	\end{table}

	\begin{figure}[h]
		\begin{center}
			\includegraphics[width=0.3\textwidth]{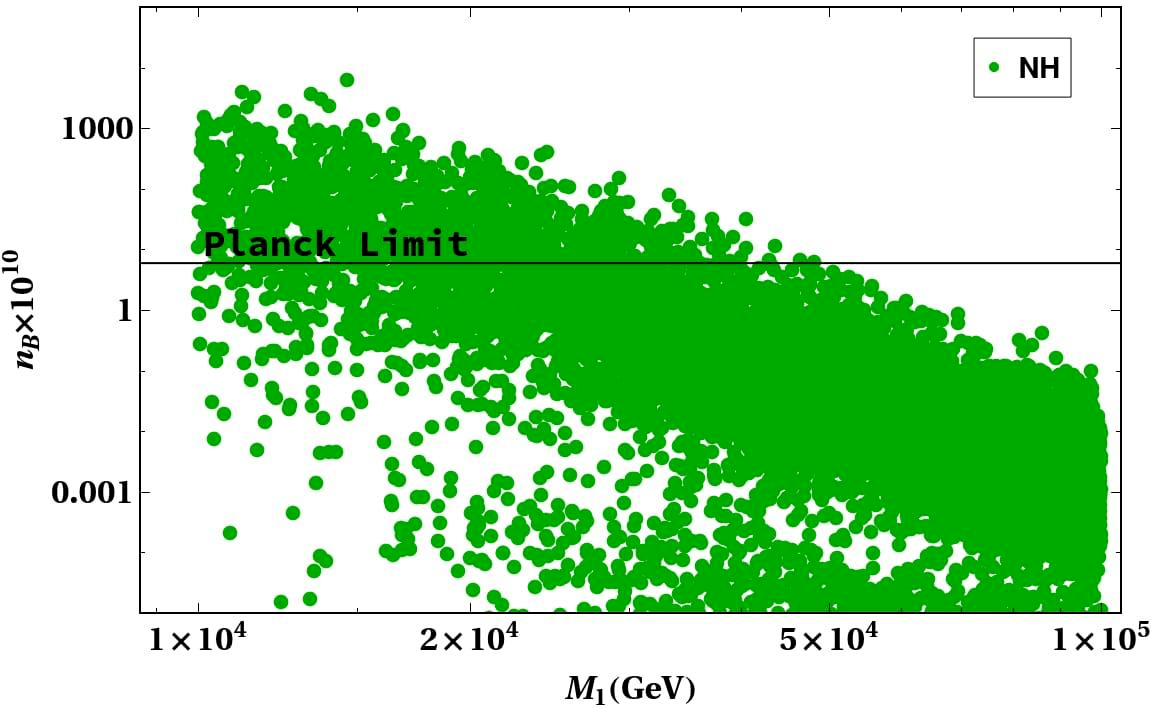}
			\includegraphics[width=0.3\textwidth]{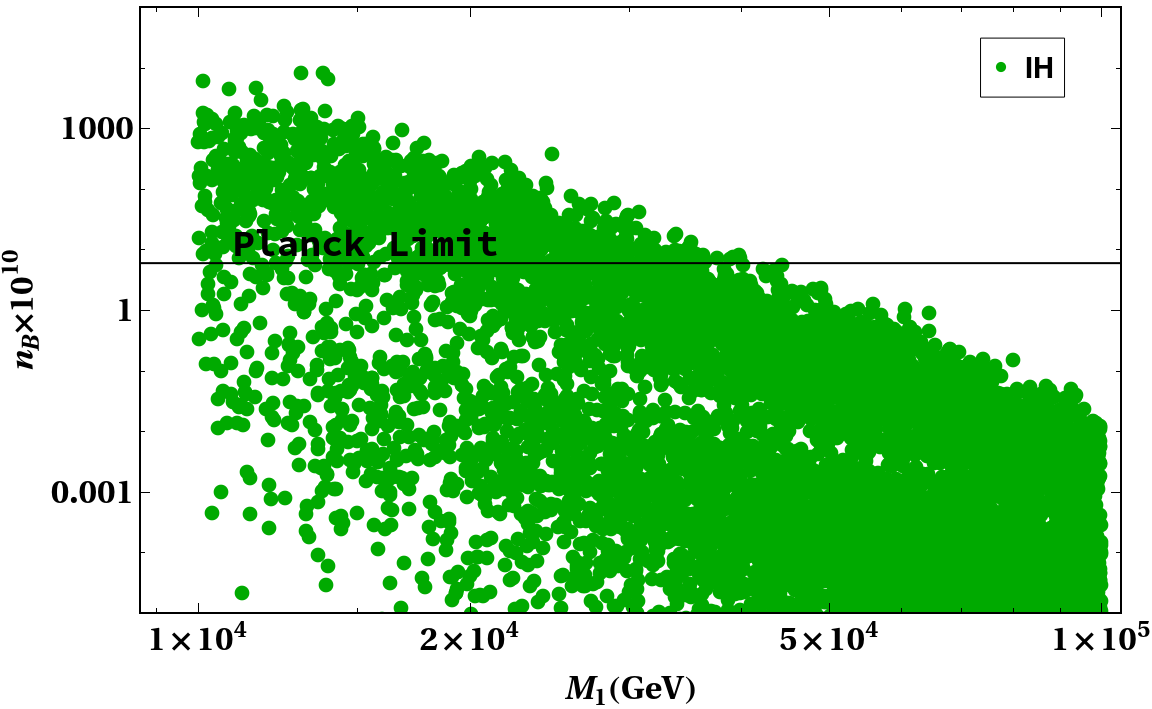}\\
			\includegraphics[width=0.3\textwidth]{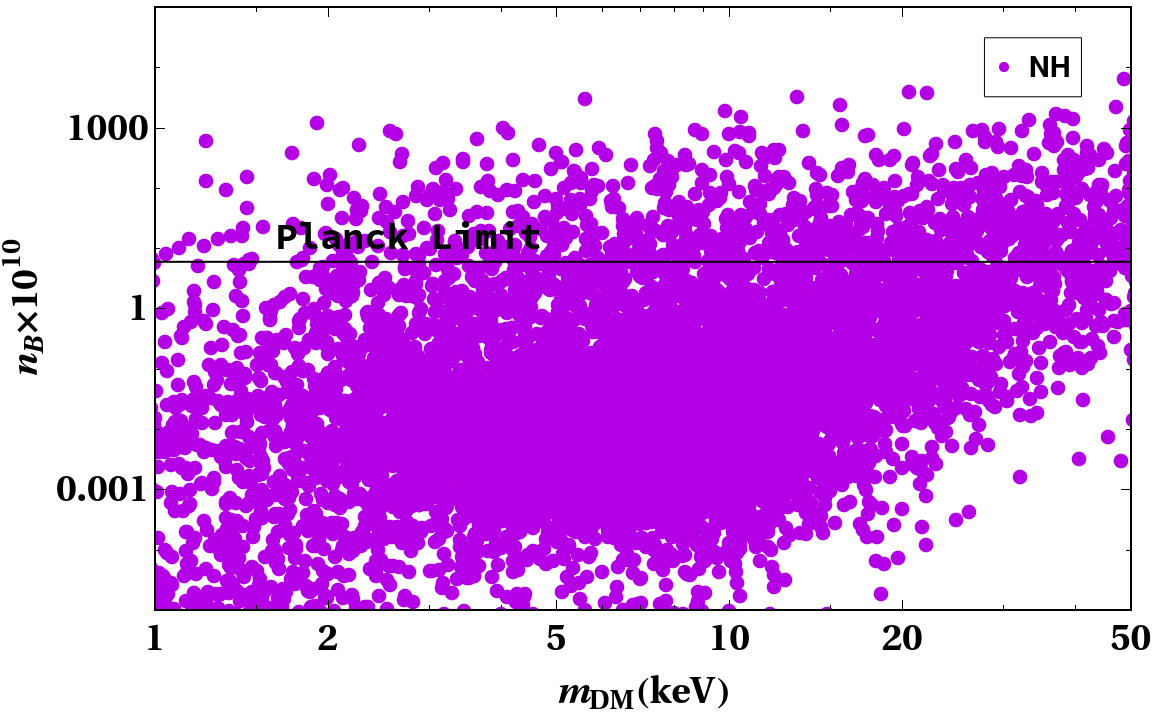}
			\includegraphics[width=0.3\textwidth]{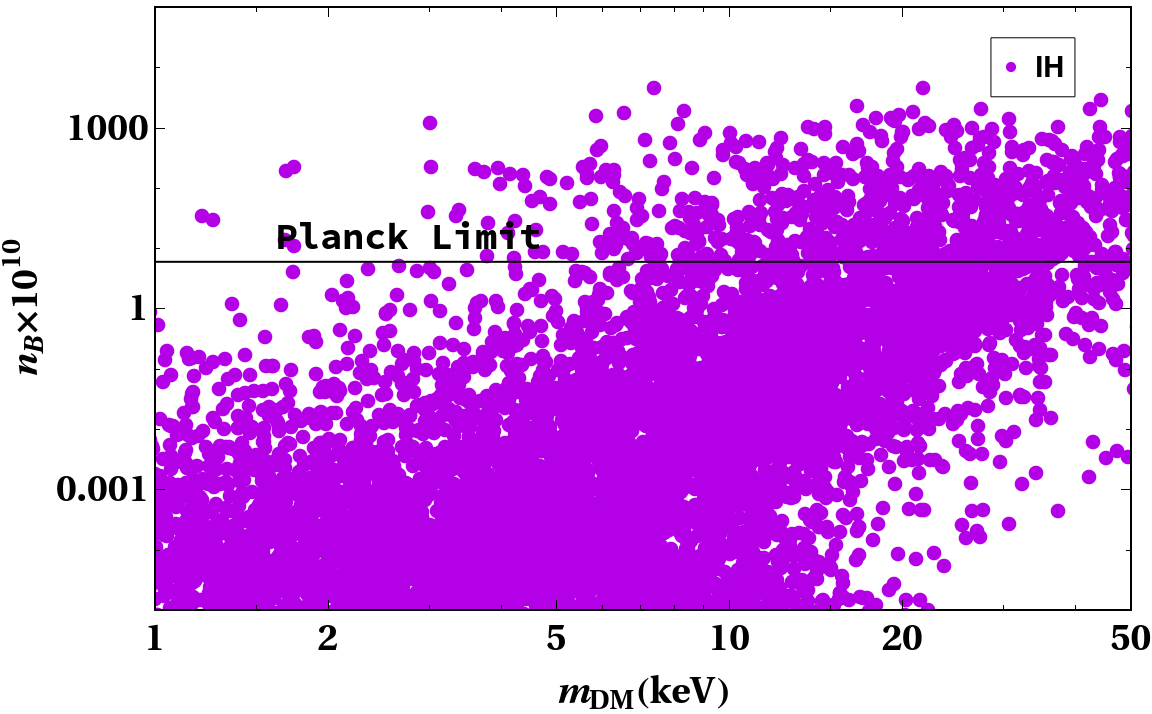}\\
			\includegraphics[width=0.3\textwidth]{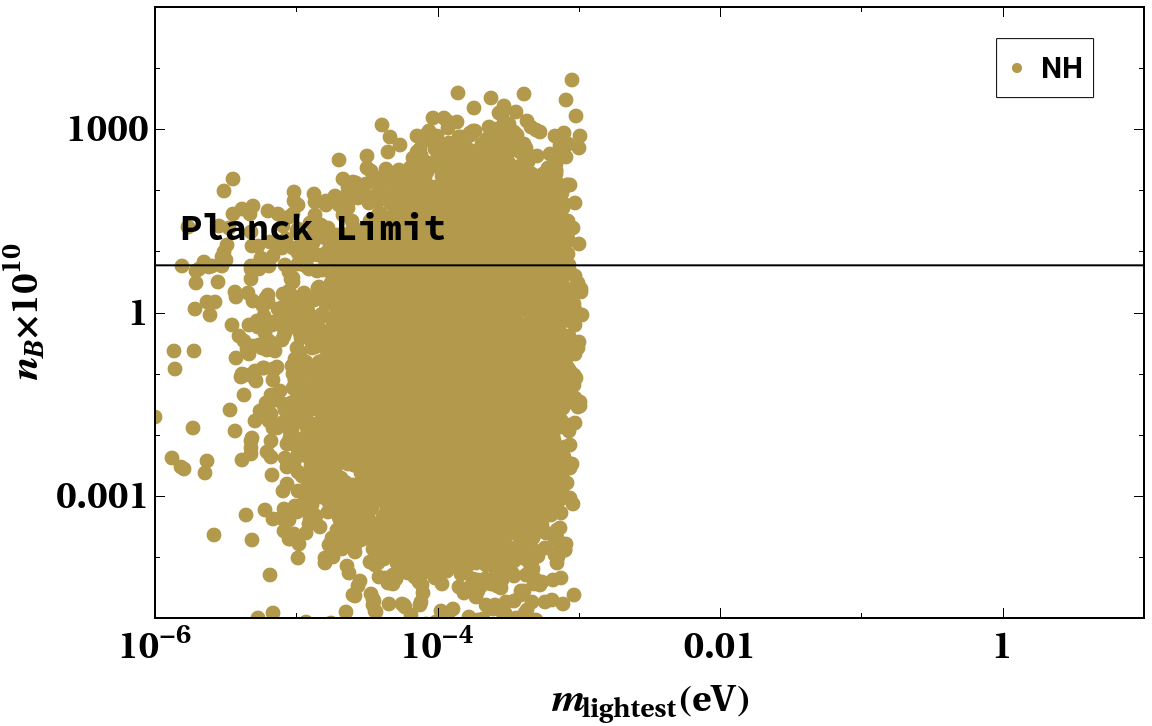}
			\includegraphics[width=0.3\textwidth]{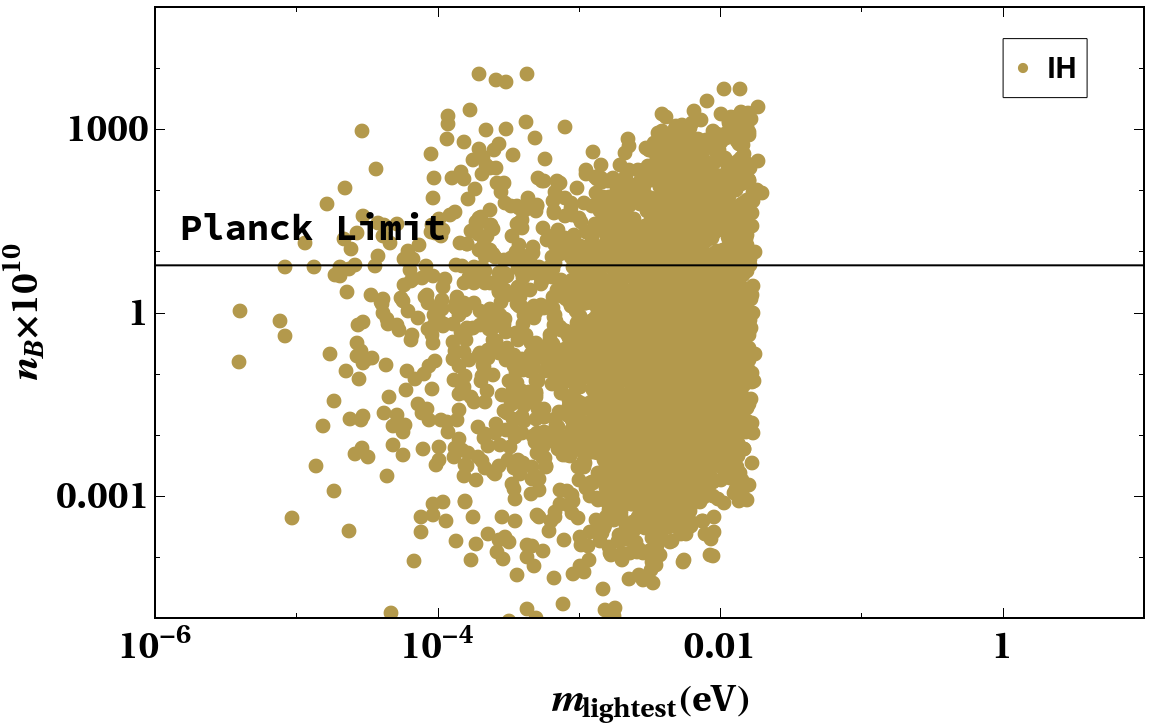}\\
			\includegraphics[width=0.3\textwidth]{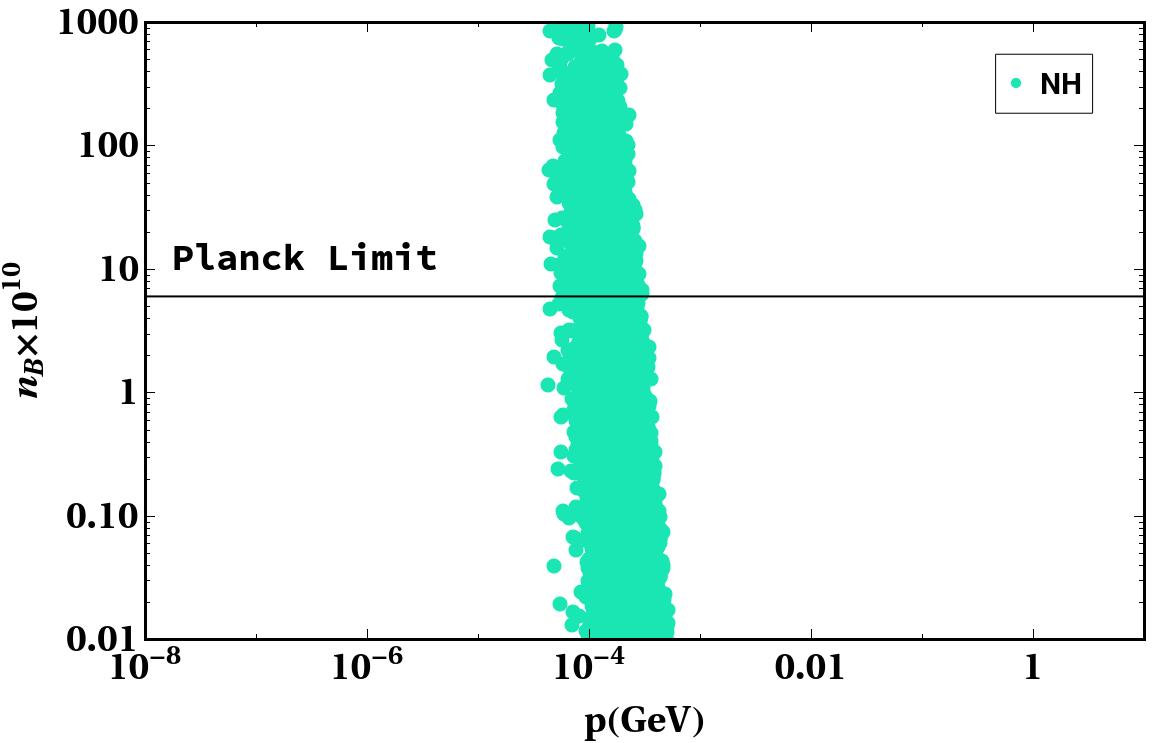}
			\includegraphics[width=0.3\textwidth]{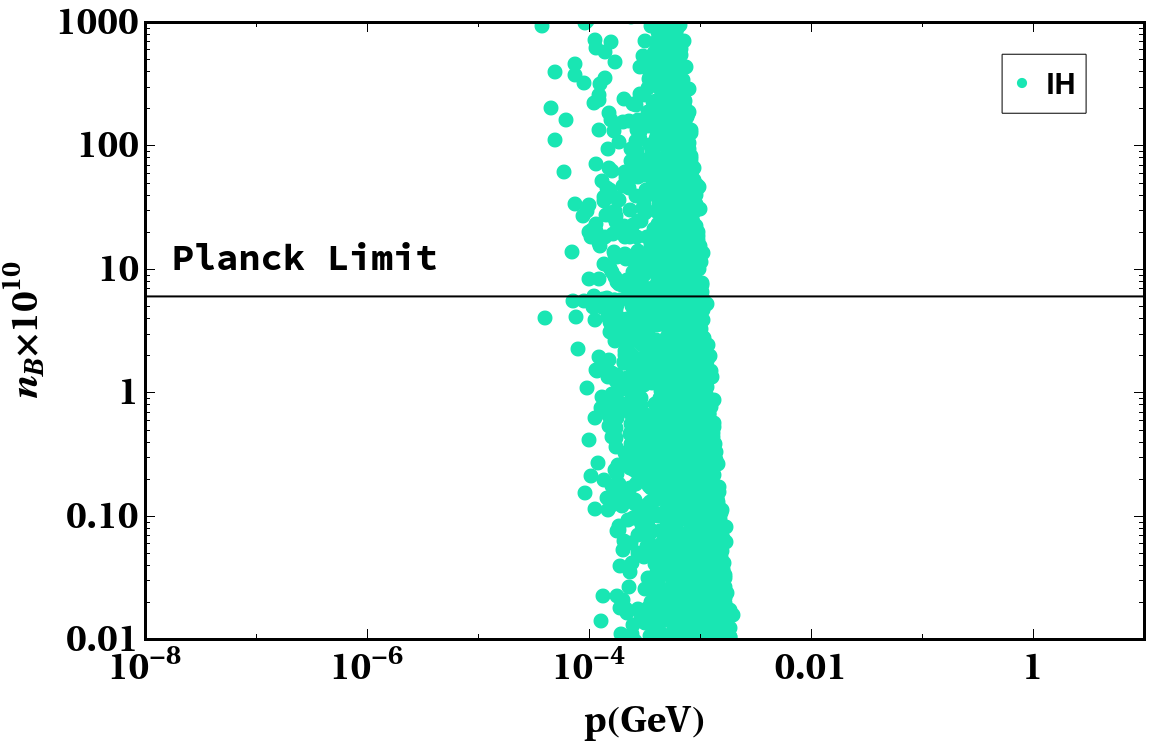}\\
			
			\caption{Plots in the first-row shows baryon asymmetry as a function of RHN mass ($M_{1}$), the second-row shows baryon asymmetry as a function of dark matter mass($M_{DM}$), in third-row baryon asymmetry as a function of lightest neutrino mass eigenvalue($m_{l}$) is depicted and in the fourth row we show a plot between perturbation(p) and BAU respectively. The black horizontal line gives the current Planck limit for BAU viz. $6.05\times10^{-10.}$. }\label{BAU}
		\end{center}
	\end{figure}
	
	\begin{figure}[h]
		\begin{center}	
			\includegraphics[width=0.45\textwidth]{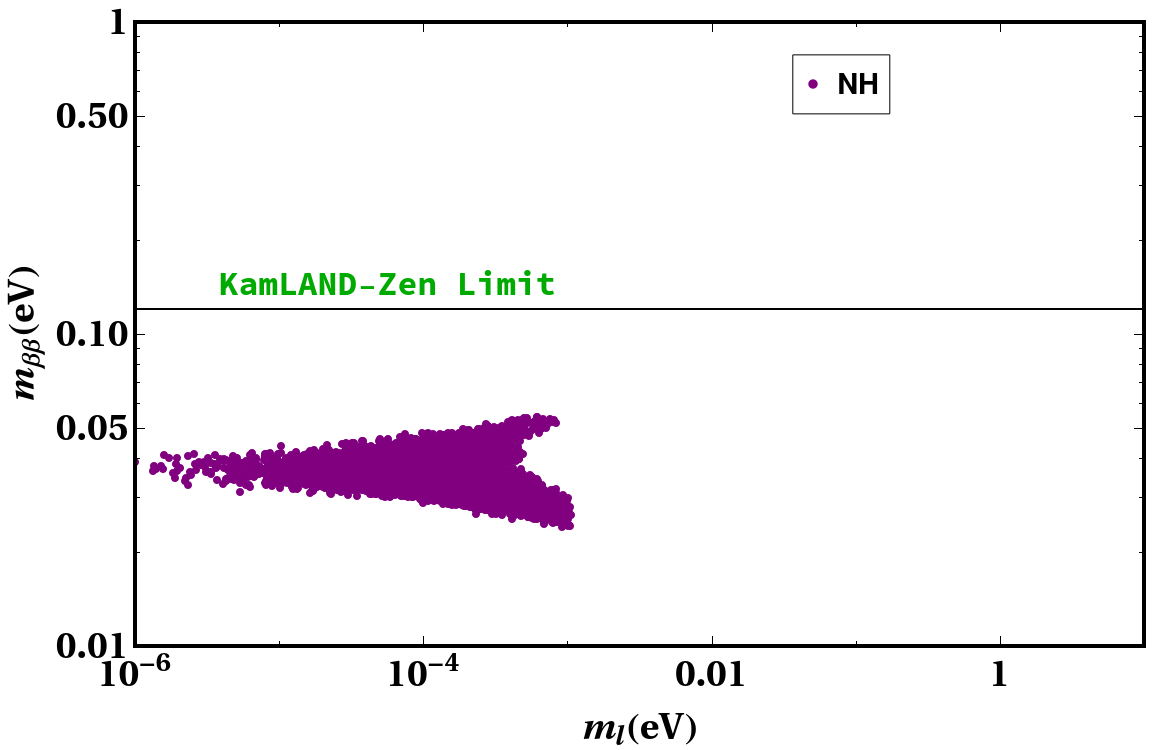}
			\includegraphics[width=0.45\textwidth]{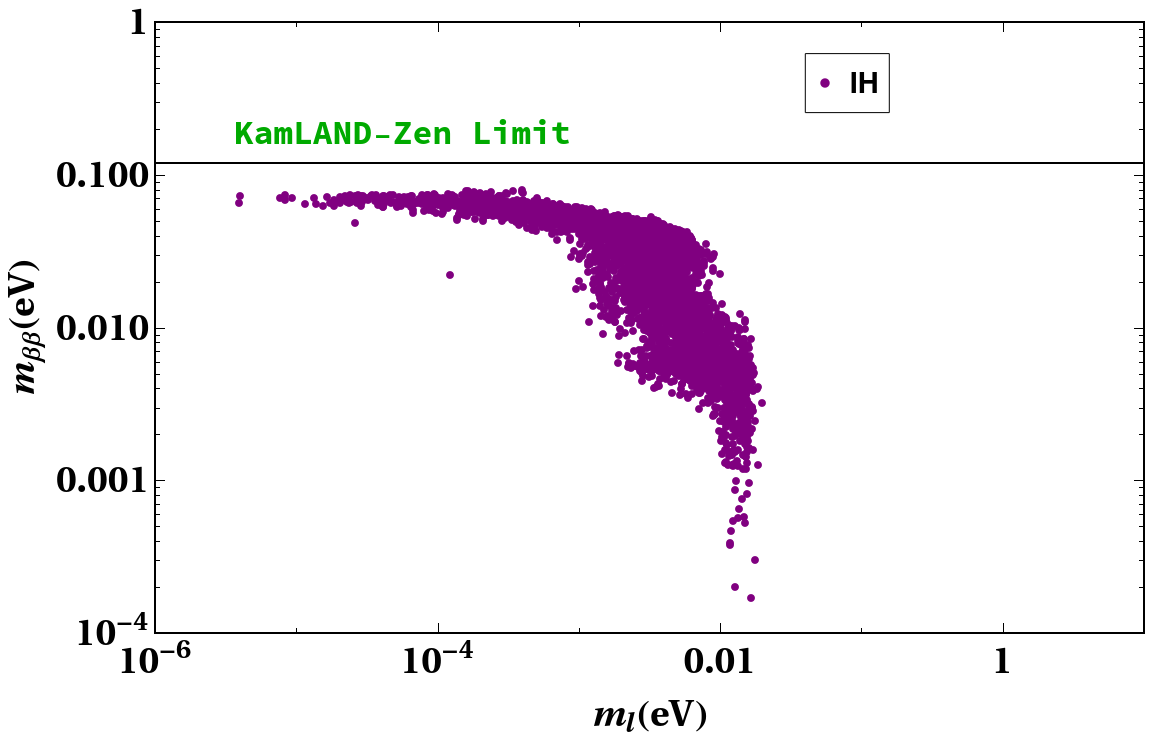}
			\caption{ Lightest active neutrino mass($m_{l}$) as a function of effective mass($m_{\beta\beta}$) for NH/IH. The horizontal(black) line depicts the upper bound on the effective mass ($m_{\beta\beta}(eV)\sim0.12(eV)$) of light neutrinos obtained from KamLAND-Zen experiment.}\label{NDBD}
		\end{center}
	\end{figure}
	\begin{figure}[h]
		\begin{center}
			\includegraphics[width=0.45\textwidth]{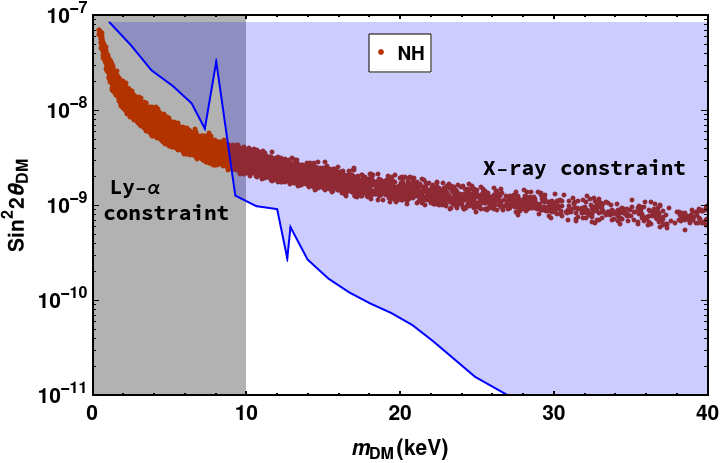}
			\includegraphics[width=0.45\textwidth]{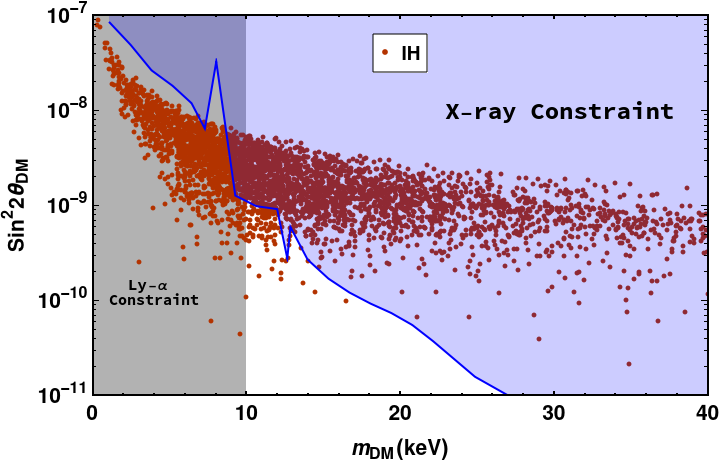}
			\caption{ Plot between dark matter mass($m_{DM}$) and active-DM mixing angle including constraints from Lyman-$\alpha$ and X-ray for NH/IH.}\label{DM}
			
		\end{center}
	\end{figure}
	
	\begin{figure}[h]
		\begin{center}
			\includegraphics[width=0.45\textwidth]{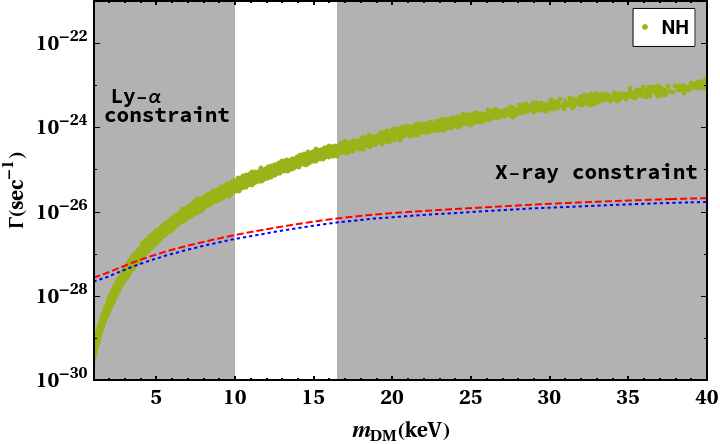}
			\includegraphics[width=0.45\textwidth]{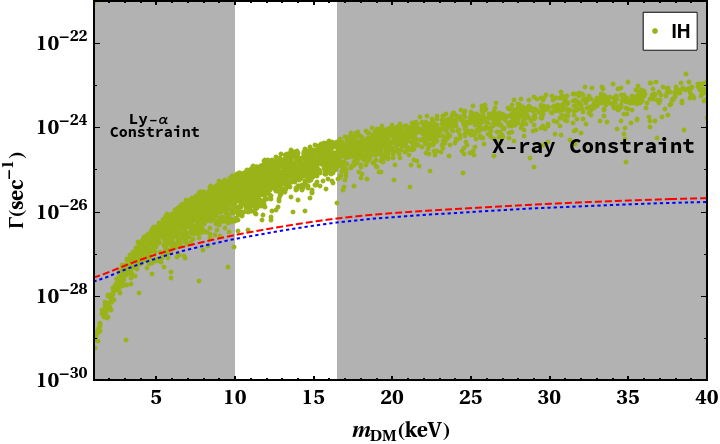}
			\caption{ Plot between dark matter mass($m_{DM}$) and decay rate for the process $S \longrightarrow \nu + \gamma $ including constraints from Lyman-$\alpha$ and X-ray for NH/IH. We also give lifetime constraint of sterile neutrinos as a function of mass. The red dashed (blue dotted) line represents the case when energy transfer from CMB photons to gas is included (excluded).} \label{DM1}
			
		\end{center}
	\end{figure}
	
	\begin{figure}[h]
		\begin{center}
			\includegraphics[width=0.45\textwidth]{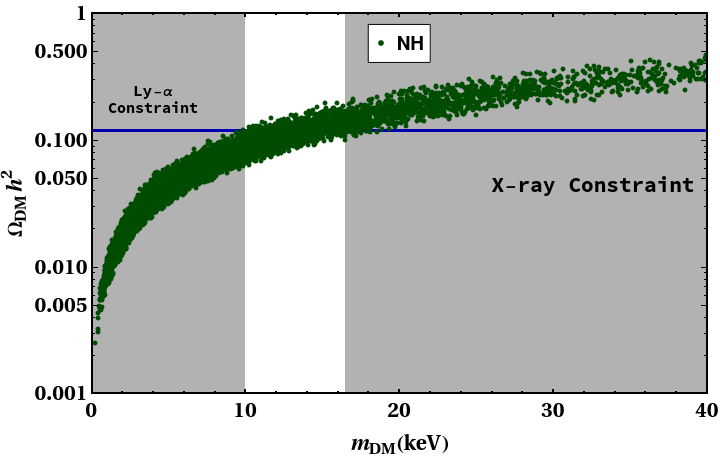}
			\includegraphics[width=0.45\textwidth]{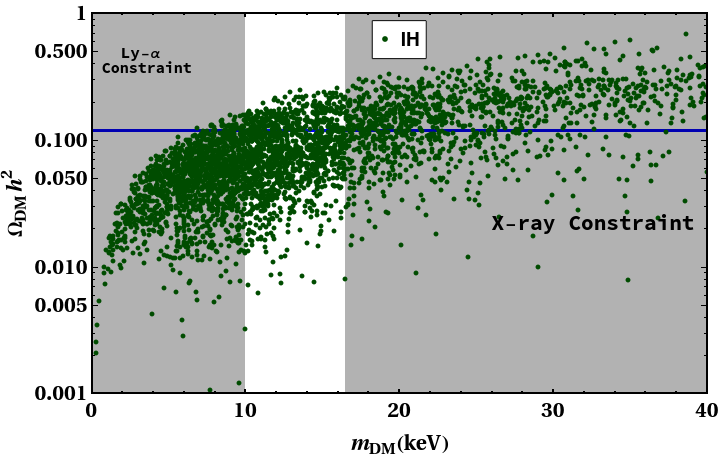}
			\caption{ Variation between dark matter mass and relic abundance with the constraints obtained from Lyman-$\alpha$ and X-ray for NH/IH. }\label{DM2}
			
		\end{center}
	\end{figure}
	
	\begin{figure}[h]
		\begin{center}
			\includegraphics[width=0.4\textwidth]{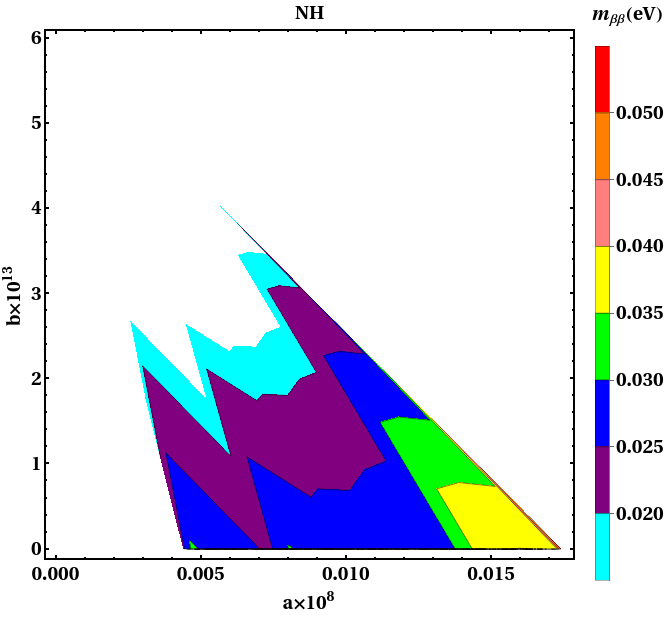}
			\includegraphics[width=0.4\textwidth]{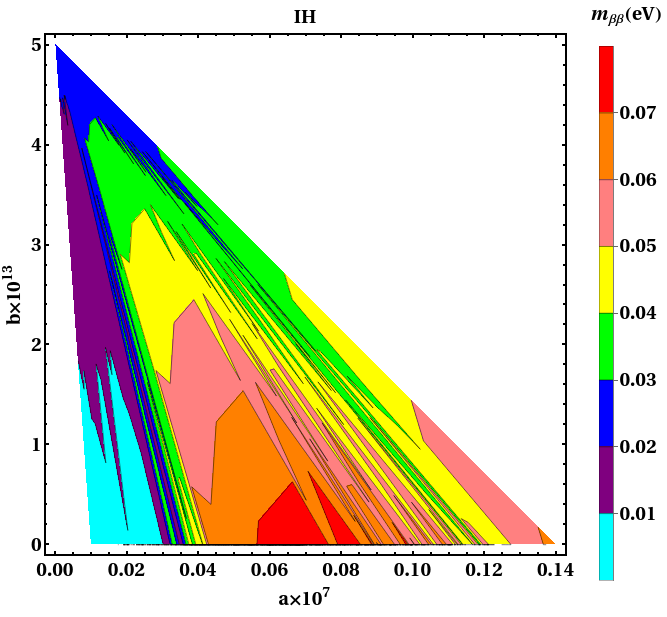}
			\caption{ Contour plot showing the parameter space of model parameters a and b w.r.t effective mass($m_{\beta\beta}$) for NH/IH. }\label{C1}
			
		\end{center}
	\end{figure}
	\begin{figure}[h]
		\begin{center}
			\includegraphics[width=0.4\textwidth]{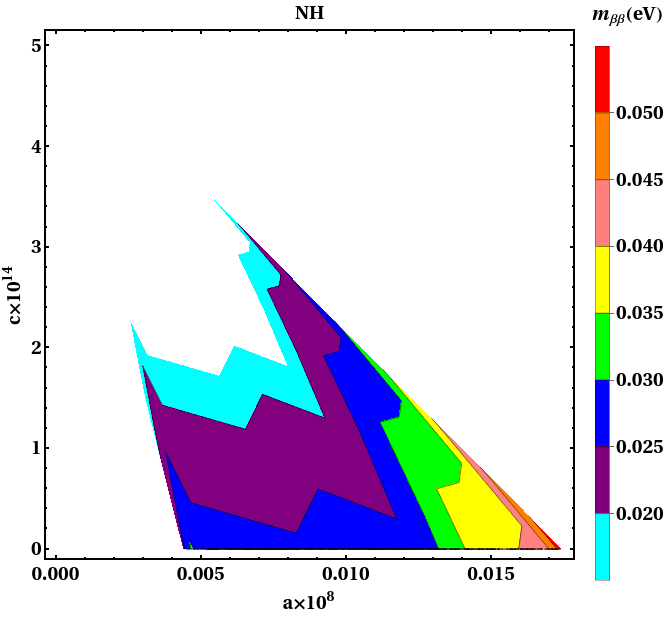}
			\includegraphics[width=0.4\textwidth]{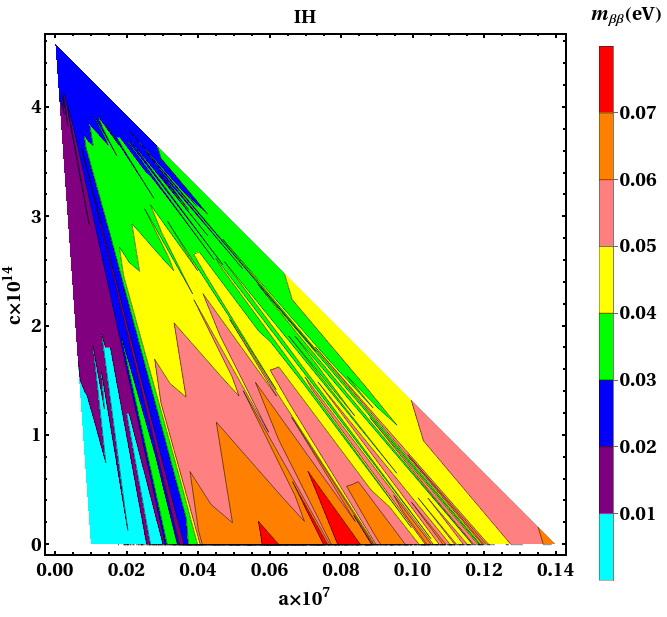}
			\caption{ Contour plot showing the parameter space of model parameters a and c w.r.t effective mass($m_{\beta\beta}$) for NH. }\label{C2}
			
		\end{center}
	\end{figure}
	
	\begin{figure}[h]
		\begin{center}
			\includegraphics[width=0.4\textwidth]{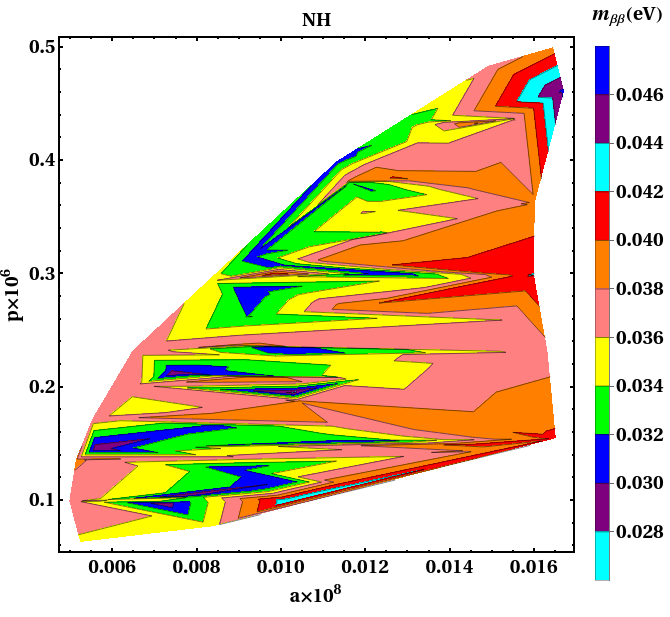}
			\includegraphics[width=0.4\textwidth]{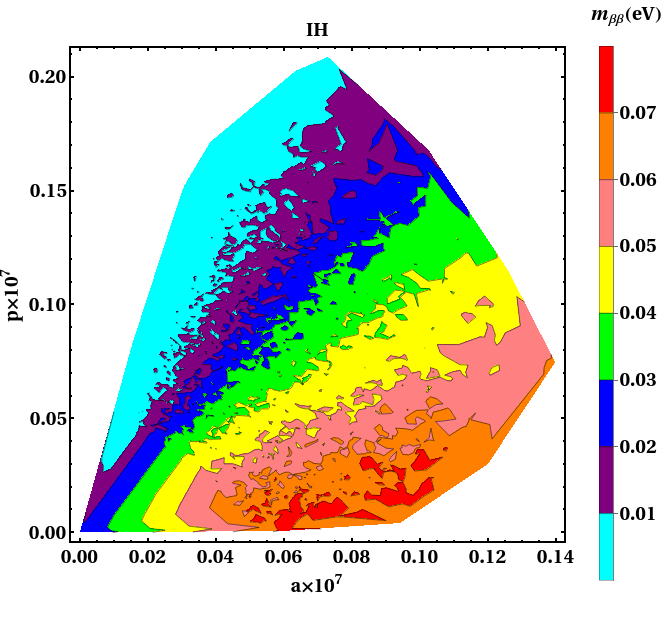}
			\caption{ Contour plot showing the parameter space of model parameters a and p w.r.t effective mass($m_{\beta\beta}$) for NH/IH. }\label{C3}
			
		\end{center}
	\end{figure}
	
	\begin{table}[h]
		\begin{center}
			\begin{tabular} { |c|c|c| }
				\hline
				Model Parameter & NH (eV) & IH (eV) \\
				\hline
				a & $0.005\times10^8$-$0.015\times10^8$ & $0.02\times10^7$- $0.14\times10^7$\\
				\hline
				b & $0.1\times10^{13}$-$3.5\times10^{13}$ & $0.1\times10^{13}$ -$5\times10^{13}$ \\
				\hline
				c & $0.1\times10^{14}$-$3\times10^{14}$ & 0.1$\times10^{14}$-$4.6\times10^{14}$ \\
				\hline
				p & $0.1\times10^{6}$-$0.5\times10^{6}$ & $0.01\times10^{7}$- $0.2\times10^{7}$ \\
				\hline
			\end{tabular}
		\end{center}
		\caption{Allowed range of the model parameters satisfying effective mass}\label{TAB3}
	\end{table}
	
	\subsection{Neutrinoless double beta decay}
	A very important open problem in neutrino physics is the search for true nature of neutrino whether its Dirac particle or Majorana particle. From the theoretical point of view it is expected that neutrinos are Majorana
	particles (the necessity of extremely small values of the neutrino Yukawa coupling constants is commonly considered as a strong argument against a SM origin of the neutrino masses). Neutrinoless double beta decay ($0\nu\beta\beta$) is one of the lepton number violating processes which can probe Majorana nature of neutrino. 
	\begin{equation}
	(A,Z)\longrightarrow (A,Z+2)+e^-+e^-
	\end{equation}
	whose amplitude is proportional to the effective Majorana mass,
	\begin{equation}
	\mid m_{\beta\beta}\mid=\mid\sum_k V^2_{ek}m_k\mid
	\end{equation}
	where V is the active-sterile mixing matrix.\\
	Using the standard parametrization of the (3+1) active-sterile mixing matrix, the effective Majorana mass in $0\nu\beta\beta$ decay can be written as
	\begin{equation}
	\mid m_{\beta\beta}\mid=\mid c_{13}^2 c_{12}^2 c_{14}^2 m_1e^{i\phi_1}+c_{13}^2 s_{12}^2 c_{14}^2m_2+s_{13}^2 c_{14}^2 m_3 e^{i\phi_2} +s_{14}^2m_4 \mid
	\end{equation}
	The parameters involved are:\\
	(i) the angles $\theta_{12}$ and $\theta_{13}$ , measured with good precision by the solar, short-baseline reactor neutrino experiments, respectively;\\
	(ii) the neutrino mass eigenstates $m_1$ , $m_2$, $m_3$ and $m_4$ , which
	are related to the solar ($\Delta m_S^2$), atmospheric ($\Delta m_A^2$) and LSND ($\Delta m_{LSND}^2$) squared mass differences :\\
	\begin{equation}
	\Delta m_S^2=\Delta m_{12}^2
	\end{equation}
	\begin{equation}
	\Delta m_A^2=\frac{1}{2}\mid\Delta m_{13}^2+\Delta m_{23}^2\mid
	\end{equation}
	\begin{equation}
	\Delta m_{LSND}^2= \Delta m_{41}^2~~ or~~ \Delta m_{43}^2
	\end{equation}
	The relation between the mass eigenstates and the squared mass differences allows two possible orderings of the neutrino masses :\\
	\textbf{Normal mass hierarchy :} ($m_1\ll m_2< m_3\ll m_4$)\\[2mm]
	$m_1=m_{min}$, $m_2=\sqrt{m_{min}^2+\Delta m_S^2}$, $m_3=\sqrt{m_{min}^2+\Delta m_A^2+\frac{\Delta m_S^2}{2}}$, $m_4=\sqrt{m_{41}^2}$\\[4mm]
	\textbf{Inverted mass hierarchy :} ($m_3\ll m_1< m_2 \ll m_4$)\\[2mm]
	$m_3=m_{min}$, $m_1=\sqrt{m_{min}^2+\Delta m_A^2-\frac{\Delta m_S^2}{2}}$, $m_2=\sqrt{m^2_{min}+\Delta m_A^2+\frac{\Delta m_S^2}{2}}$, $m_4=\sqrt{m_{43}^2}$

	\subsection{Results}
	A cumulative study has been carried out in this flavor symmetric $\nu$2HDM so as to have a wider perspective on the viability of the model. The choice of free parameters considered in our work is given in Table \ref{TAB2}. The entire work is carried out keeping these benchmark parameter space fixed. From Fig.\eqref{BAU}, we obtain variational plots between baryon asymmetry of the Universe calculated from the model and the free parameters considered in our work except for the plot in the last row. We have chosen a diagonal perturbation matrix in order to break the $\mu-\tau$ symmetry of the neutrino mass matrix, thus, in last row of Fig.\eqref{BAU} we show the parameter space of perturbation($p$) satisfying the Planck limit for BAU. We find only a narrow region of space ranging from $10^{-4}-10^{-3}$ GeV which have points obeying constraints for BAU in case of both NH/IH. Also the range of RHN mass corresponding to Planck limit for both the hierarchies are same, i.e. $10^{4}-5\times 10^{4}$ GeV as shown in first row of Fig.\eqref{BAU}. However, there is a difference in the allowed parameter space for dark matter mass. For NH, the entire DM mass range $1-50$ keV satisfies the BAU bound whereas in case of IH, we have distinct allowed points in the range $5-50$ keV. The lightest active neutrino mass is also constrained in our work, wherein the allowed range for NH is approximately $10^{-5}-10^{-3}$ eV and that for IH is $10^{-4}-10^{-2}$ eV. In Fig.\eqref{NDBD}, both NH and IH satisfies the KamLAND-Zen limit for effective mass in variation with $m_{l}$. As discussed in sec.\eqref{sec3}, the Lyman-$\alpha$ bound on $m_{DM}$ forbids mass below $10$ keV in case of non resonantly production of sterile neutrino (which serve as a DM candidate in our case). Results of dark matter phenomenology is discussed in the points below:\\
		$\bullet$ Fig.\eqref{DM} shows a co-relation between dark matter mass and active-DM mixing angle. A very small region between $10-16$ keV falls in the allowed parameter space for IH whereas the entire parameter space for NH falls in the excluded region.\\
		$\bullet$ We have also shown a plot of dark matter mass w.r.t the decay rate ($\Gamma$) for both NH and IH respectively in Fig.\ref{DM1}. The allowed space consists of points corresponding to dark matter mass $10-16$ keV in context with the bounds coming from Lyman-$\alpha$ and X-ray for both NH and IH. Though we have shown the variation plot of $m_{DM}$ vs. $\Gamma$, but the mass region for NH is not allowed as obtained from Fig.\ref{DM}. We have also taken into account constraints from lifetime of sterile neutrinos w.r.t its mass. As can be seen from Fig.\ref{DM1}, the red dashed line corresponds to the upper bound coming from lifetime of sterile neutrinos for the case when energy transfer from cosmic microwave background(CMB) photons to gas is included and the blue dotted lines is for the case when this very energy is excluded. Considering this constraint, the results obtained excludes all the points for NH, whereas for IH, we have very scanty points in the boundary of $10$ keV which tends to satisfy the lifetime bounds.\\
		$\bullet$ Similarly in Fig.\ref{DM2}, we have shown the dark matter mass range which obeys the Planck limit for relic abundance of dark matter. In case of NH, $m_{DM}=10-16$ keV satisfies the stringent bound for relic abundance though this region is disallowed from Fig.\ref{DM}. We have larger number of points for IH in the allowed parameter space $m_{DM}$=$10-16$ keV satisfying the Planck bound.\\
	Fig.\eqref{C1},\eqref{C2} and \eqref{C3} depicts the range of the model parameters a, b, c and the perturbation p consistent with the KamLAND-Zen limit for effective mass of neutrinos. A tabular form comprising of the allowed space of the model parameters is depicted in Table.\eqref{TAB3}.\\
	We cannot comment on the more preferable hierarchy for neutrino phenomenology or baryon asymmetry of the Universe due to their identical allowed parameter space. However, in case of dark matter we can consider IH to have shown a wider range of parameter space when compared to NH which abide by the experimental or observational constants.
	
	\section{CONCLUSION} \label{sec6}
	In our work we have studied neutrino phenomenology and related cosmology of an extension of $\nu$2HDM realized with the help of $A_4 \otimes Z_8$ flavor symmetries. The particle content of our model includes three right handed neutrino fields ($N_{1}, N_{2}, N_{3}$), one Higgs doublet ($\eta$), one additional gauge singlet ($S$) and four sets of flavon fields ($\varphi, \xi, \chi, \zeta$) to the Standard Model of particle physics. The Dirac mass matrix ($M_D^\prime$), Majorana mass matrix ($M_R$), sterile mass matrix ($M_S$) are constructed as required using the $A_4$ product rules. A perturbation ($M_P$) is incorporated in the $M_D^\prime$ in order to break the $\mu-\tau$ symmetry in the light neutrino mass matrix to generate the non zero reactor mixing angle and obtain a $4\times4$ active-sterile mass matrix similar to the MES framework. We analyze both the normal and inverted hierarchies extensively in this work. The model parameters are solved comparing the active-sterile neutrino matrix diagonalized by the active-sterile mixing matrix in Eq.\eqref{eq:7}. After evaluating the model parameters, the sterile neutrino mass(which is the Dark Matter candidate) and DM-active mixing angle is calculated. In this work, we have considered non resonant production of sterile neutrino and thus the constraints from Lyman-$\alpha$ and X-ray are implemented accordingly. The variation of active-DM mixing angle with DM mass is studied and it is observed that the data points satisfy the Lyman-$\alpha$ and X-ray constraints in the IH case but are disfavored for NH. The decay rate of the DM for the process $S \longrightarrow \nu + \gamma $ is also calculated. We obtain a low decay rate which establishes the stability of the dark matter candidate in the cosmological scales. The relic abundance of the dark matter candidate is also checked and studied w.r.t the variation in DM mass. Also in this case, we see that the number of data points that satisfy the Lyman-$\alpha$ and X-ray constraints in IH are more than that compared to NH. Overall, the dark matter phenomenology is more compatible for IH. Baryogenesis is also studied in our work due to its crucial role in phenomenological analysis. In our model, BAU is generated through the out of equilibrium decay of $N_{1}\rightarrow l\eta, \bar{l}\eta^{*}$, where $N_1$ is lightest right handed neutrino. We study baryon asymmetry as a function of lightest right handed neutrino mass ($N_1$), DM mass, lightest active neutrino mass and perturbation ($p$) considering constraints from Planck limit. We also calculate the effective neutrino mass and then study it's variation with lightest active neutrino mass and validate with KamLAND-Zen limit. The allowed parameter space of model parameters is generated w.r.t the effective mass of active neutrinos. In conclusion we can say that IH has shown a wider range of parameter space in case of dark matter compared to NH. An identical allowed parameter space in the both hierarchies is seen in neutrino phenomenology and BAU. Thus, we can consider our model to be consistent in addressing dark matter, neutrino phenomenology and baryon asymmetry of the universe simultaneously.
	
\bibliographystyle{JHEP}

\providecommand{\href}[2]{#2}\begingroup\raggedright\endgroup

\end{document}